\documentclass[reprint,superscriptaddress,amsmath,amssymb,floatfix]{revtex4-2}

\usepackage{graphicx}
\usepackage{bm}
\usepackage[colorlinks=true,linkcolor=blue,urlcolor=blue,citecolor=blue,pdfstartview=FitH]{hyperref}
\usepackage{amsmath}
\usepackage{microtype}
\usepackage{siunitx}
\usepackage{txfonts}

\usepackage{xcolor}
\usepackage{soul}

\usepackage[normalem]{ulem}

\newcommand{\ju}[1]{\textcolor{black}{{#1}}}

\newcommand{\su}[1]{\textcolor{black}{{#1}}}
\newcommand{\vd}[1]{\textcolor{black}{{#1}}}

\begin{document}

\title{Competing routes to spontaneous flow in confined active nematics}

\author{Rahil N. Valani}
\affiliation{Rudolf Peierls Centre for Theoretical Physics, Parks Road,
University of Oxford, OX1 3PU, United Kingdom}
\email{rahil.valani@physics.ox.ac.uk\\ julia.yeomans@physics.ox.ac.uk}

\author{Vedad Dzanic}
\affiliation{School of Mechanical, Medical, and Process Engineering,
Queensland University of Technology, Brisbane, QLD 4001, Australia}
\email{v2.dzanic@qut.edu.au}

\author{Julia M. Yeomans}
\affiliation{Rudolf Peierls Centre for Theoretical Physics, Parks Road,
University of Oxford, OX1 3PU, United Kingdom}

\author{Sumesh P. Thampi}
\affiliation{Department of Chemical Engineering,
Indian Institute of Technology Madras, Chennai 600036, India}
\email{sumesh@iitm.ac.in}


\date{\today}

\begin{abstract}
Active nematics can spontaneously develop flow beyond a critical activity in confined geometries. We show analytically that the onset of flow is governed by two competing instabilities: a long-wavelength mode leading to unidirectional flow and a finite-wavelength instability producing transverse rolls. A reduced description reveals how activity, flow alignment, and nematic elasticity control the competition between these modes. We identify regimes in which the finite-wavelength instability has a lower critical activity than the long-wavelength instability, causing vortices to emerge before unidirectional flow. This establishes wavelength selection as an intrinsic feature of the onset of spontaneous flow in active nematics.
\end{abstract}

\maketitle

\textit{Introduction --}
Active nematics are orientationally ordered systems whose constituents continuously convert energy into mechanical stresses. Examples range from suspensions of microtubules driven by kinesin motors to dense bacterial systems and layers of elongated biological cells \cite{Doostmohammadi2018,Sanchez2012,Doostmohammadi2022,Gompper2020}. Unlike passive nematic liquid crystals, active nematics generate flows internally. Distortions of the orientational order produce active stresses that drive fluid motion, which in turn reorients the nematic. This feedback underlies a rich variety of collective dynamics, including spontaneous flows, coherent vortices, motile topological defects, and active turbulence \cite{Wensink2012,Shendruk2017}. Understanding how an ordered quiescent state first loses stability is central to understanding how these complex dynamical states emerge.

\begin{figure}
    \centering
    \includegraphics[width=1\linewidth]{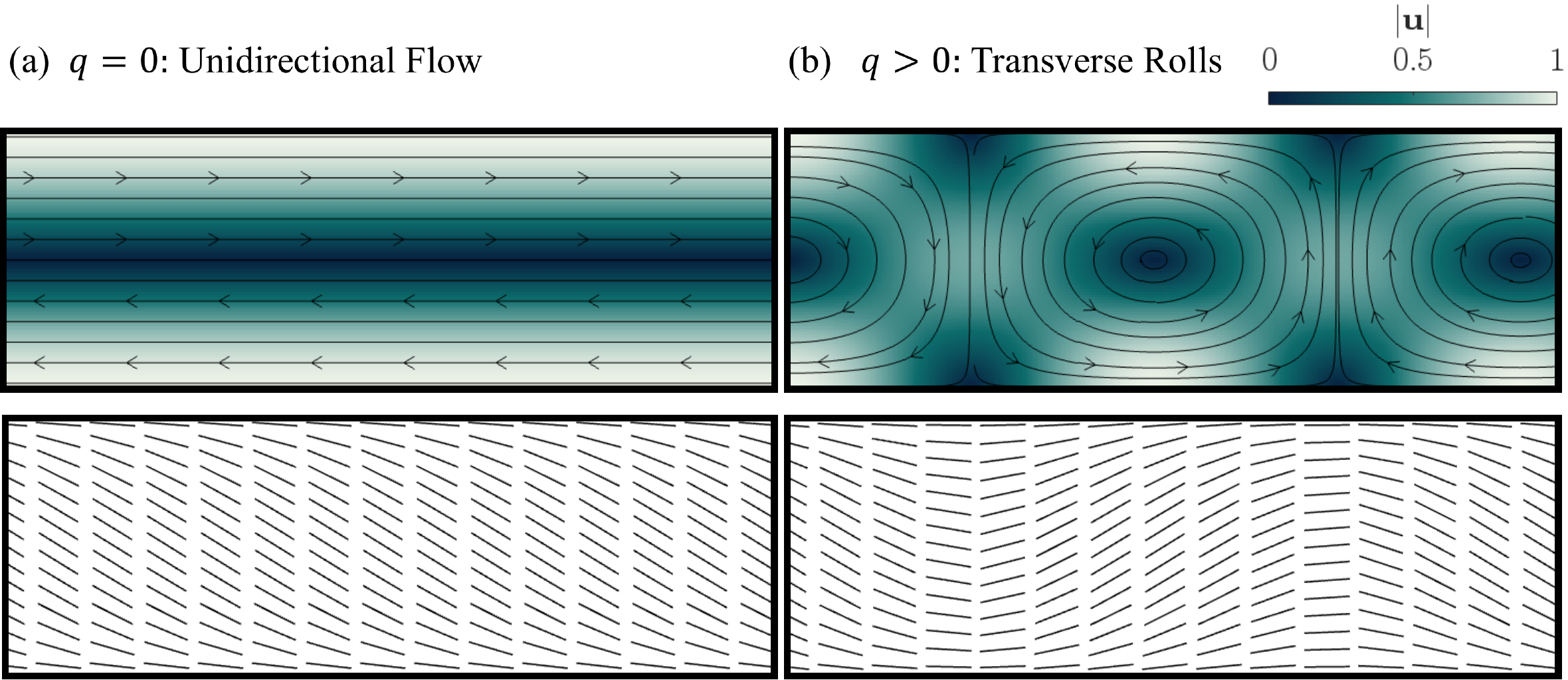}
\caption{Schematic of the two competing modes at the onset of spontaneous flow in a confined geometry:
(a) long-wavelength ($q=0$) mode, corresponding to
unidirectional flow, (b) finite-wavelength
($q>0$) transverse roll mode. Top panels show the velocity field, with
background color indicating the velocity magnitude $|\mathbf{u}|$ and superimposed streamlines indicating the flow direction. Bottom panels show the corresponding nematic
configuration, with line segments representing the local director orientation. 
}
    \label{fig:schematic}
\end{figure}

A fundamental result of active-nematic hydrodynamics is that activity can destabilize uniform orientational order through the coupling between director distortions and the flows they generate \cite{Ramaswamy2002,Voituriez2005}. 
In an unbounded system, this instability occurs at long wavelengths, rendering the ordered state generically unstable.
Confinement suppresses these modes and introduces a finite threshold for spontaneous flow \cite{Voituriez2005}. 
In channel geometries, the primary instability is conventionally associated with the emergence of unidirectional flow (Fig.~\ref{fig:schematic}(a)), with its onset and stability controlled by activity, confinement, and the response of the nematic director to shear \cite{Thampi2022}.

Beyond this primary transition, confined active nematics support a hierarchy of increasingly complex flow states. With increasing activity, unidirectional flow can destabilize into oscillatory and vortical states  and, at stronger driving, active turbulence \cite{Shendruk2017, Opathalage2019, Hardouin2019, Samui2021, sousa2026spontaneous}. Finite-wavelength structures have therefore largely been viewed as secondary instabilities that develop after spontaneous flow has already emerged. However, 
this raises a fundamental question: can a finite-wavelength mode become unstable before the conventional long-wavelength 
mode and thereby become the primary route to spontaneous flow\su{s and active turbulence}?

Here, we answer this question analytically. We derive a reduced Galerkin description of the overdamped Beris–Edwards equations \cite{zienkiewicz2013,Beris1994} and obtain an explicit growth rate for the linear modes of the quiescent confined state. The resulting dispersion relation reveals a competition between activity, nematic elasticity, and flow alignment that determines not only the instability threshold but also the wave number of the fastest-growing mode. Crucially, the analytic result predicts a regime in which the leading unstable mode occurs at a finite wave number, with transverse rolls (Fig.~\ref{fig:schematic}(b)) emerging as the primary instability of the quiescent state. Our results\su{, also validated via full numerical simulations,} establish wavelength selection as an intrinsic part of the spontaneous-flow transition and provide a simple framework for determining not only when an active nematic begins to flow, but which flow state emerges first.


\begin{figure}
    \centering
    \includegraphics[width=1\linewidth]{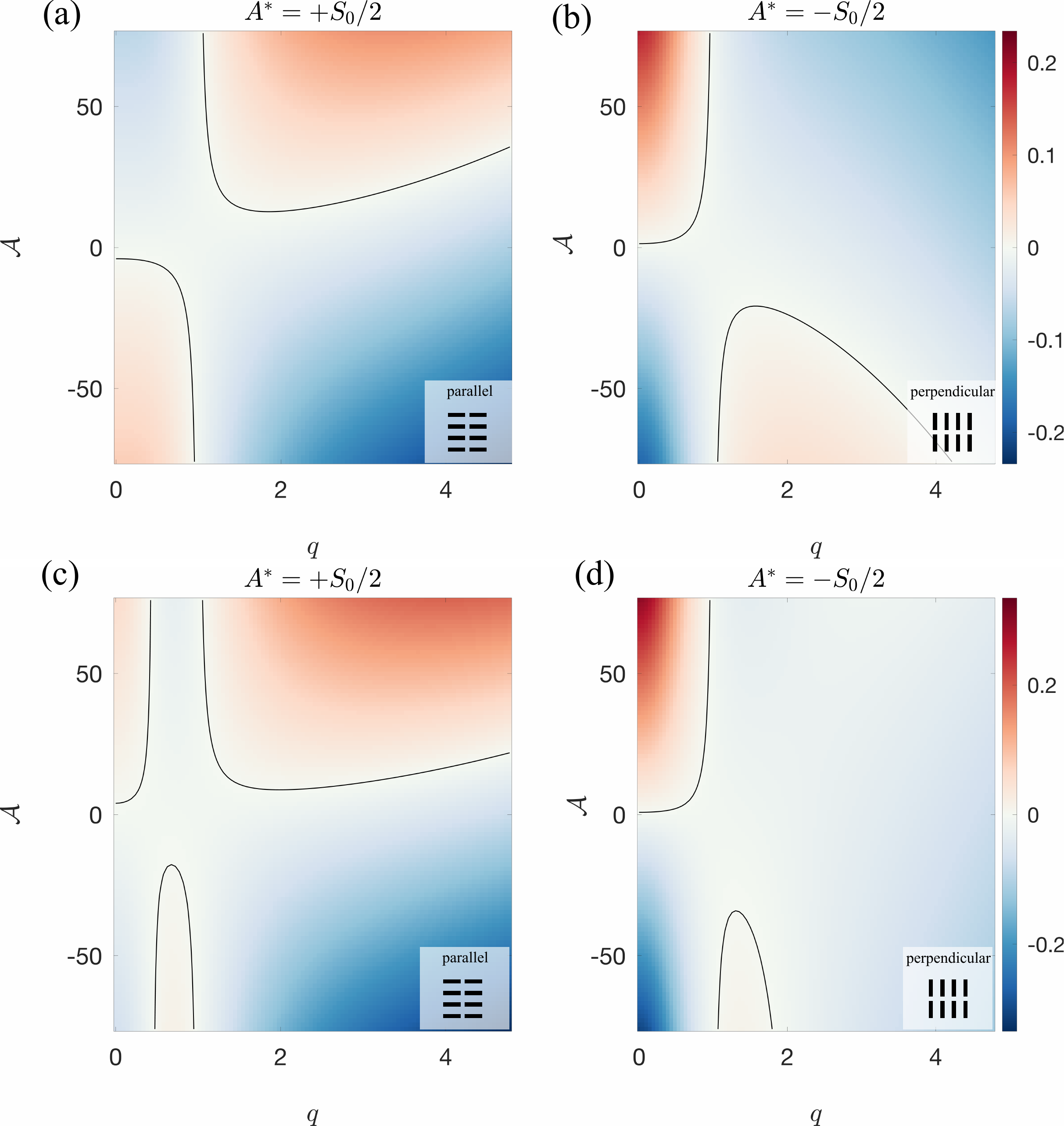}
\caption{
Linear growth rate $\mu$ of transverse roll perturbations to the quiescent active-nematic state in the activity--wavenumber plane
\su{($\mathcal{A},q)$}. Left and right columns correspond to instability of uniformly ordered base states
with the director parallel ($A^*=+S_0/2$) and perpendicular
($A^*=-S_0/2$) to the $x$-direction, respectively. Panels (a,b) show
$\lambda=0.5$, while (c,d) show $\lambda=1.5$. The color scale gives the
growth rate, with $\mu>0$ ($\mu<0$) indicating unstable (stable) transverse
perturbations. Black contours mark the neutral-stability boundary $\mu=0$.
The remaining parameters are \su{$K = 0.0031, C = 0.032$},
$S_0=1$, and $k_0=2\pi/L_0$, with $L_0=20$. }
    \label{fig:F1}
\end{figure}


\begin{figure*}
    \centering
\includegraphics[width=0.7\linewidth]{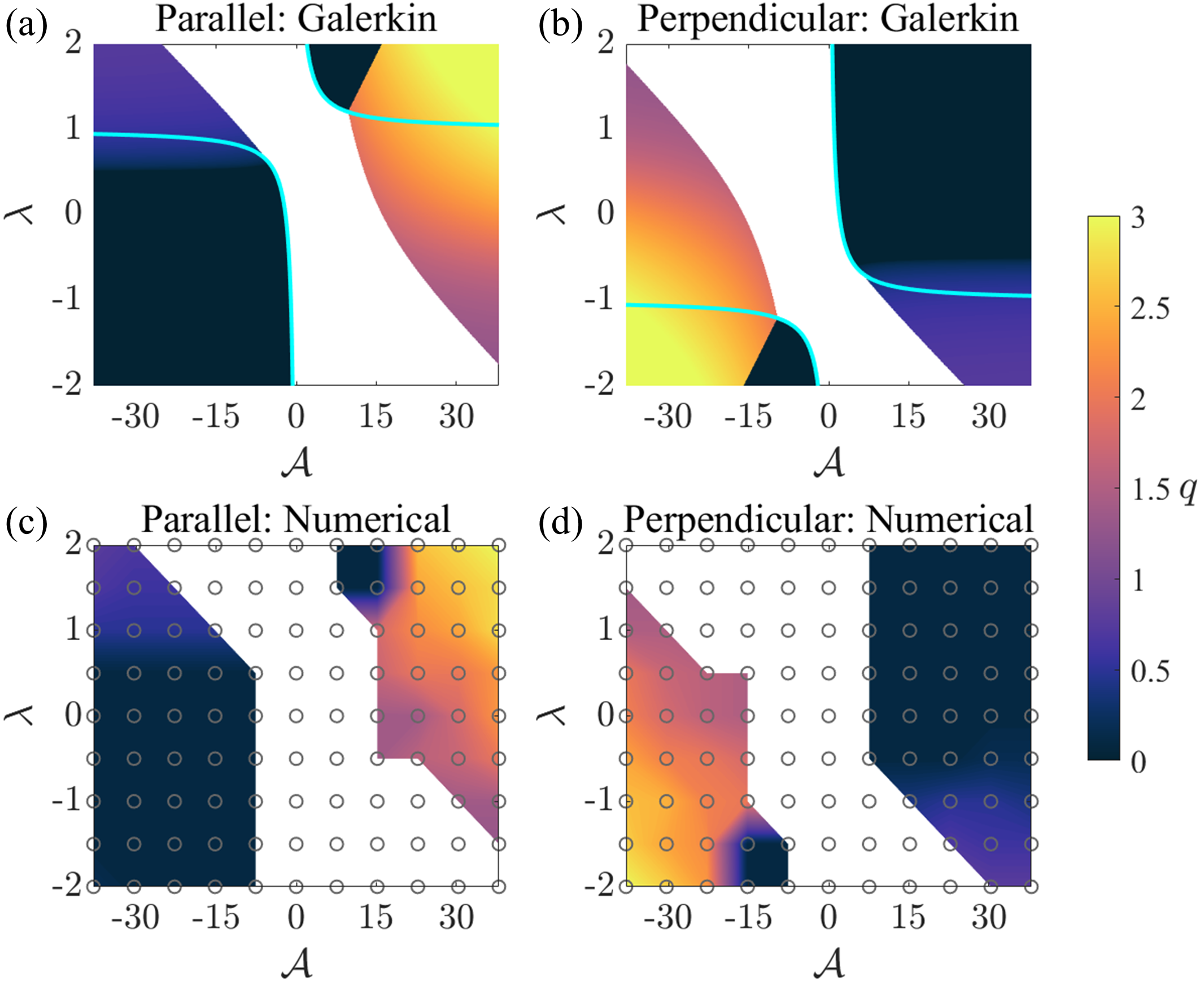}
\caption{
Most unstable transverse wavenumber, \su{$q$},  in the activity \su{number}--flow-alignment plane
\su{$(\mathcal{A},\lambda)$}. Panels (a) and (b) show results from the linear stability analysis for perturbations to uniformly
ordered states with the director parallel ($A^*=+S_0/2$) and perpendicular
($A^*=-S_0/2$) to the $x$-direction, respectively. (c) and (d) are from numerical solutions of the equations of motion \eqref{eq:stokes}-\eqref{eq:BE}. The color scale denotes
the wavenumber of the fastest-growing mode: dark regions correspond to the
long-wavelength instability, whereas nonzero wavenumbers indicate selection
of a finite-wavelength transverse roll mode. The cyan curves mark the
one-dimensional stability boundaries, highlighting regions where the
finite-wavelength instability can pre-empt the conventional unidirectional
route to spontaneous flow. All other parameters are as in
Fig.~\ref{fig:F1}.}
    \label{fig:F2}
\end{figure*}

\textit{Model --}
We consider a two-dimensional active nematic fluid governed by the Beris--Edwards
equations \cite{Beris1994}. In the creeping flow limit, the non-dimensional equations describing the evolution of the velocity field $\mathbf{u}$, the incompressibility criterion, and the evolution of the nematic order parameter tensor, $\mathbf{Q}$, are
\begin{align}
0=-\nabla p+\nabla^2\mathbf{u}
-\alpha\nabla\cdot\mathbf{Q},
\qquad
\nabla\cdot\mathbf{u}=0,
\label{eq:stokes}\\
\partial_t\mathbf{Q}
+\mathbf{u}\cdot\nabla\mathbf{Q}
-\mathbf{S}
=
K\nabla^2\mathbf{Q}
+C\left(S_0^2-2\,\mathrm{Tr}\,\mathbf{Q}^2\right)\mathbf{Q}
\label{eq:BE}
\end{align}
where $\mathbf{Q} =\begin{pmatrix}
Q_{xx} & Q_{xy}\\
Q_{xy} & -Q_{xx}
\end{pmatrix}$  is a symmetric, second rank tensor.

In Eq.~\eqref{eq:stokes}, $p$ is the pressure. The activity coefficient $\alpha$ drives the system out of equilibrium. $\alpha > 0$ corresponds to extensile and $\alpha < 0$ to contractile nematic activity.

In Eq.~\eqref{eq:BE}, $S_0$ sets the preferred magnitude of the nematic order parameter at equilibrium and $C$ controls the relaxation of $\mathbf{Q}$, while $K$ is an elastic coefficient, penalizing gradients in $\mathbf{Q}$. 
$\mathbf{S}$ is the generalised co-rotational derivative,
\begin{align}
\mathbf{S}={}&
(\lambda\mathbf{E}+\mathbf{\Omega}) \cdot
\left(\mathbf{Q}+\frac12\mathbf{I}\right)
+\left(\mathbf{Q}+\frac12\mathbf{I}\right)
\cdot (\lambda\mathbf{E}-\mathbf{\Omega})
\nonumber\\
&-2\lambda
\left(\mathbf{Q}+\frac12\mathbf{I}\right)
(\mathbf{Q}:\mathbf{E}),
\label{eq:S_dim}
\end{align}
where $\lambda$ is the flow-alignment parameter, and
$\mathbf{E}=[\nabla\mathbf{u}+(\nabla\mathbf{u})^T]/2$ and
$\mathbf{\Omega}=[\nabla\mathbf{u}^T-(\nabla\mathbf{u})]/2$ are the strain rate and vorticity tensors\vd{,} respectively.

To non-dimensionalise the equations, the following characteristic scales were assumed: $k_0^{-1}$ for length, $U_0$ for velocity and  $\eta k_0 U_0$ for pressure where, $\eta$ is the shear viscosity of the fluid. \su{The strength of activity compared to elasticity is quantified by defining the activity number $\mathcal{A} = \alpha/K$.} 

Considering an active nematic in the domain $-\infty<x<\infty$, $-\pi/2<y<\pi/2$, we first identify, as base states for the linear stability analysis, the homogeneous, quiescent states,
\begin{align}
\mathbf{u} = 0; \quad Q_{xx} = A^* = \pm S_0/2 \quad \textnormal{and} \quad Q_{xy} = 0.
\label{eq:base}
\end{align}
The two ordered branches correspond to director fields parallel
($A^*=+S_0/2$) or perpendicular ($A^*=-S_0/2$) to  $x$. They will be referred to as, respectively, the `parallel' and `perpendicular' quiescent states. 




For further analysis, we use the strategy of Lorenz's low-dimensional reduction by retaining a minimal set of Fourier modes. Following a procedure similar to that used in Rayleigh--B\'enard convection \cite{Lorenz1963,Das2020}, we perturb each ordered state with a minimal roll mode 
$f(x,y)=\cos(qx)\cos y$,
\begin{equation}
\psi=X(t)f(x,y),\quad \textnormal{and} \quad
\delta Q_{xy}=Y(t)f(x,y)
\label{eq:ansatz}
\end{equation}
while holding $Q_{xx}=A^*$. Here, $\psi$ is the stream function: $\mathbf{u} = (\partial_y\psi) \hat{x} - (\partial_x\psi) \hat{y} $ and $q$ is the non-dimensional wavenumber in the $x$-direction.

Defining $\Lambda=q^2 + 1$ and $\Delta=q^2-1$, taking the curl of Eq.~\eqref{eq:stokes} and projecting onto $f$, we obtain
\begin{equation}
X=\frac{\alpha\Delta}{\Lambda^2}Y.
\label{eq:slaving}
\end{equation}
The perturbations in the flow field are, therefore, instantaneously slaved to the perturbative distortions in the nematic order, reflecting the absence of fluid inertia in Eq.~(\ref{eq:stokes}).


Projecting the linearized Beris--Edwards equation onto the same mode and
using Eq.~\eqref{eq:slaving} we obtain
\begin{equation}
\dot Y=\mu(q)Y,
\label{eq:ydot}
\end{equation}
where
\begin{equation}
\mu(q)=
\frac{\alpha\Delta}{\Lambda^2}
\left(A^*\Lambda+\frac{\lambda}{2}\Delta\right)
-K\Lambda
+C\left(S_0^2-4A^{*2}\right).
\label{eq:mu}
\end{equation}
Equation~\eqref{eq:mu} is the central result of the reduced model. It determines both the onset of the transverse instability and the preferential amplification of the spatial scales. The detailed Galerkin projection leading to Eq.~\eqref{eq:mu} is provided in the End Matter.


\color{black}

\textit{Results --}
We first examine the stability of the ordered quiescent states to the roll
mode. A perturbation grows when $\mu(q)>0$, with neutral stability defined by $\mu(q)=0$. For $A^*=\pm S_0/2$, the criterion for neutral stability is
\begin{equation}
\frac{\alpha(q^2-1)}{(q^2+1)^2}
\left(
A^*(q^2+1)+\frac{\lambda}{2}(q^2-1)
\right)
=K(q^2+1) .
\label{eq:thresholdordered}
\end{equation}
The threshold activity for the instability reflects a competition between active amplification and elastic
relaxation. An orientational distortion generates an active flow that can,
through flow alignment, reinforce the distortion, while elasticity
increasingly damps spatial variations at larger wavenumber. Because the
active feedback itself depends on $q$, this competition can select a finite
spatial scale at the instability.

We note that the limiting case of roll perturbations, \textit{i.e.,} when $q \rightarrow 0$ is the well-known conventional, long wavelength one-dimensional instability of active nematics \cite{Ramaswamy2002, Voituriez2005} (Fig.~\ref{fig:schematic}(a)). For an extensile (contractile) system, this long wavelength instability produces bend (splay) distortions in the ordered director field. On the other hand, the growth of a finite-wavenumber mode, $q>0$, gives rise to the roll modes shown in Fig.~\ref{fig:schematic} (b).

Fig.~\ref{fig:F1} shows the growth rate of the unstable modes in the activity--wavenumber plane for the parallel and perpendicular quiescent states. The black contours are neutral stability curves ($\mu=0$) that delimit the regions unstable to roll perturbations.   The response of the system depends strongly on both the initial director orientation and the flow alignment parameter. For $\lambda=0.5$ (Fig.~\ref{fig:F1}(a)), in the flow-tumbling regime, the parallel quiescent state is unstable at $q \sim 0$ for contractile activity and at finite $q$ for extensile activity. The opposite is true for the perpendicular quiescent state (Fig.~\ref{fig:F1}(b)). Thus, long-wavelength instabilities are favored for contractile (extensile) activity in the parallel (perpendicular) quiescent state.
Increasing the flow-alignment parameter to $\lambda=1.5$, corresponding to the flow aligning regime, qualitatively reorganizes the
stability landscape (Fig.~\ref{fig:F1}(c)-(d)). Most notably, for both the parallel and perpendicular quiescent states, a region of instability at $q=0$ is allowed for extensile, but not for contractile, driving.  
Therefore, flow alignment changes not only the threshold activity required for instability, but also the range of spatial scales that can be amplified.

An unstable finite-wavelength mode does not, in itself, imply that rolls are always selected as the quiescent state becomes unstable. The roll instability must dominate the one-dimensional instability ($q = 0$) when both are present. We, therefore, maximize $\mu(q)$ over the allowed wavenumbers for each activity and flow-alignment parameter to identify the most unstable mode.

Fig.~\ref{fig:F2} summarizes the resulting mode selection in the \su{$\lambda$ - $\mathcal{A}$} plane. Panels (a) and (b) respectively show the most unstable wavenumber of the parallel and perpendicular states. Over wide range of parameters in the \su{$\lambda$ - $\mathcal{A}$} plane, roll modes (non-zero $q$)  can replace the conventional, one-dimensional instability as the fastest growing mode. The plots also clearly respect the symmetry of the active nematic equations of motion under the mapping $\lambda \Leftrightarrow -\lambda$, \su{$\mathcal{A} \Leftrightarrow -\mathcal{A} $}, $Q \Leftrightarrow -Q $ (director rotated by $\pi/2$).  

Moreover, Fig.~\ref{fig:F2} shows that the quiescent state can be unstable to roll modes in regions of the $\lambda$--\su{$\mathcal{A}$} plane where it is predicted to be stable in the $q=0$ limit by the one-dimensional linear stability analysis \cite{Ramaswamy2002}. \su{Two-dimensional numerical simulations, however, show instabilities and active turbulence in these regions, which cannot be explained by the one-dimensional stability analysis.} Our two-dimensional stability analysis resolves this discrepancy by demonstrating that the roll modes span an instability basin, of which the long-wavelength instability region forms only a subset.

In these regions, the usual sequence of flow states, \textit{i.e.,} $\text{quiescent}\rightarrow\text{unidirectional flow}\rightarrow\text{spatially structured flow}$ is preempted, and the system can instead undergo a direct transition from $\text{quiescent}\rightarrow\text{transverse rolls}$. Thus, the roll state is not necessarily a secondary instability \cite{Samui2021, sousa2026spontaneous} of an existing flow, but can constitute a distinct primary linear instability of the ordered quiescent nematic. Moreover, when extended to confined geometries, such as a channel, our analysis shows that the preferred mode is not determined solely by the confinement scale, but also varies with activity and flow alignment.


Next, we test the predictions of our roll mode stability analysis against numerical solutions of the full governing equations of motion. We use a hybrid lattice Boltzmann technique for the computer simulations, and thus even inertial terms of the Navier Stokes equations are included in the calculations. The fastest growing modes at early times observed in the simulations are in excellent agreement with the analytical predictions [Fig.~\ref{fig:F2}(c)-(d)], thus both justifying the minimal Galerkin ansatz used in this analysis and also validating the predicted instability boundaries from the roll mode analysis.  We find that simulations initialized with a random director field can evolve toward either unidirectional or roll flows, indicating possible multistability of the system. 

\su{Both analytical and numerical results presented in Fig.~\ref{fig:F2} indicate that the unstable wave numbers associated with roll modes range primarily in $q \lesssim 1 $ to $q \gtrsim 1 $. In other words, the roll modes need not be isotropic; they span the confinement width in the $y-$ direction and a length slightly larger or smaller than the confinement width in the $x-$ direction. Indeed such rectangular rolls with aspect ratio $q $ $\lesssim 1$ or $\gtrsim 1$ are usually observed in both experiments and computer simulations \cite{Hardouin2019, Shendruk2017, Samui2021}.}

\textit{Discussion and Conclusions --}
We employ an analytical linear stability analysis based on a minimal set of Fourier modes to show that spontaneous roll flows in an  active nematic can emerge through primary instabilities. A finite-wavelength transverse mode can pre-empt the conventional instability creating a unidirectional flow, allowing rolls to emerge directly from the quiescent state rather than as a secondary instability of an existing flow. Extending to confined systems such as channels, the onset of spontaneous flow is therefore not determined by a threshold activity alone, but also involves the selection of the symmetry and characteristic length scale of the emerging state. Flow alignment plays a central role in this selection by controlling the competition between long- and finite-wavelength modes.

More broadly, our results connect spontaneous flow in active nematics to pattern selection in nonequilibrium systems. 
\su{Extending this picture to three dimensions may also shed light on the transitions between roll-like and unidirectional flow states underlying the large-scale coherent flows observed in 3D channels \cite{wu2017transition} as the channel aspect ratio is varied. Whether the exponential vortex-area statistics \cite{giomi2015geometry} reported in collective epithelial flows \cite{blanch2018turbulent} also arise from this roll-mediated route, and what this might reveal about the connection between instability mechanism and emergent flow statistics of active turbulence, remains an intriguing question for future study.}

 Our results therefore shift the question of spontaneous-flow onset from simply `\emph{when does an active nematic begin to flow?}' to `\emph{which flow state is selected when it does?}' A direct transition to rolls provides a different starting point for the development of vortical and chaotic flows than a transition through a unidirectional state, potentially offering distinct routes toward active turbulence. 



\textit{Acknowledgements -- } \ju{RV acknowledges the support of the Leverhulme Trust [Grant No. LIP-2020-014]. RV and JMY acknowledge the ERC Advanced Grant ActBio (funded as UKRI Frontier Research Grant EP/Y033981/1). SPT acknowledges funding from the Department of Science and Technology, India, under research grant CRG/2023/000169.}

\bibliography{references}

\begin{thebibliography}{20}%
\makeatletter
\providecommand \@ifxundefined [1]{%
 \@ifx{#1\undefined}
}%
\providecommand \@ifnum [1]{%
 \ifnum #1\expandafter \@firstoftwo
 \else \expandafter \@secondoftwo
 \fi
}%
\providecommand \@ifx [1]{%
 \ifx #1\expandafter \@firstoftwo
 \else \expandafter \@secondoftwo
 \fi
}%
\providecommand \natexlab [1]{#1}%
\providecommand \enquote  [1]{``#1''}%
\providecommand \bibnamefont  [1]{#1}%
\providecommand \bibfnamefont [1]{#1}%
\providecommand \citenamefont [1]{#1}%
\providecommand \href@noop [0]{\@secondoftwo}%
\providecommand \href [0]{\begingroup \@sanitize@url \@href}%
\providecommand \@href[1]{\@@startlink{#1}\@@href}%
\providecommand \@@href[1]{\endgroup#1\@@endlink}%
\providecommand \@sanitize@url [0]{\catcode `\\12\catcode `\$12\catcode `\&12\catcode `\#12\catcode `\^12\catcode `\_12\catcode `\%12\relax}%
\providecommand \@@startlink[1]{}%
\providecommand \@@endlink[0]{}%
\providecommand \url  [0]{\begingroup\@sanitize@url \@url }%
\providecommand \@url [1]{\endgroup\@href {#1}{\urlprefix }}%
\providecommand \urlprefix  [0]{URL }%
\providecommand \Eprint [0]{\href }%
\providecommand \doibase [0]{https://doi.org/}%
\providecommand \selectlanguage [0]{\@gobble}%
\providecommand \bibinfo  [0]{\@secondoftwo}%
\providecommand \bibfield  [0]{\@secondoftwo}%
\providecommand \translation [1]{[#1]}%
\providecommand \BibitemOpen [0]{}%
\providecommand \bibitemStop [0]{}%
\providecommand \bibitemNoStop [0]{.\EOS\space}%
\providecommand \EOS [0]{\spacefactor3000\relax}%
\providecommand \BibitemShut  [1]{\csname bibitem#1\endcsname}%
\let\auto@bib@innerbib\@empty
\bibitem [{\citenamefont {Doostmohammadi}\ \emph {et~al.}(2018)\citenamefont {Doostmohammadi}, \citenamefont {Ign{\'e}s-Mullol}, \citenamefont {Yeomans},\ and\ \citenamefont {Sagu{\'e}s}}]{Doostmohammadi2018}%
  \BibitemOpen
  \bibfield  {author} {\bibinfo {author} {\bibfnamefont {A.}~\bibnamefont {Doostmohammadi}}, \bibinfo {author} {\bibfnamefont {J.}~\bibnamefont {Ign{\'e}s-Mullol}}, \bibinfo {author} {\bibfnamefont {J.~M.}\ \bibnamefont {Yeomans}},\ and\ \bibinfo {author} {\bibfnamefont {F.}~\bibnamefont {Sagu{\'e}s}},\ }\bibfield  {title} {\bibinfo {title} {Active nematics},\ }\href@noop {} {\bibfield  {journal} {\bibinfo  {journal} {Nature Communications}\ }\textbf {\bibinfo {volume} {9}},\ \bibinfo {pages} {3246} (\bibinfo {year} {2018})}\BibitemShut {NoStop}%
\bibitem [{\citenamefont {Sanchez}\ \emph {et~al.}(2012)\citenamefont {Sanchez}, \citenamefont {Chen}, \citenamefont {DeCamp}, \citenamefont {Heymann},\ and\ \citenamefont {Dogic}}]{Sanchez2012}%
  \BibitemOpen
  \bibfield  {author} {\bibinfo {author} {\bibfnamefont {T.}~\bibnamefont {Sanchez}}, \bibinfo {author} {\bibfnamefont {D.~T.~N.}\ \bibnamefont {Chen}}, \bibinfo {author} {\bibfnamefont {S.~J.}\ \bibnamefont {DeCamp}}, \bibinfo {author} {\bibfnamefont {M.}~\bibnamefont {Heymann}},\ and\ \bibinfo {author} {\bibfnamefont {Z.}~\bibnamefont {Dogic}},\ }\bibfield  {title} {\bibinfo {title} {Spontaneous motion in hierarchically assembled active matter},\ }\href@noop {} {\bibfield  {journal} {\bibinfo  {journal} {Nature}\ }\textbf {\bibinfo {volume} {491}},\ \bibinfo {pages} {431} (\bibinfo {year} {2012})}\BibitemShut {NoStop}%
\bibitem [{\citenamefont {Doostmohammadi}\ and\ \citenamefont {Ladoux}(2022)}]{Doostmohammadi2022}%
  \BibitemOpen
  \bibfield  {author} {\bibinfo {author} {\bibfnamefont {A.}~\bibnamefont {Doostmohammadi}}\ and\ \bibinfo {author} {\bibfnamefont {B.}~\bibnamefont {Ladoux}},\ }\bibfield  {title} {\bibinfo {title} {Physics of liquid crystals in cell biology},\ }\href@noop {} {\bibfield  {journal} {\bibinfo  {journal} {Trends in Cell Biology}\ }\textbf {\bibinfo {volume} {32}},\ \bibinfo {pages} {140} (\bibinfo {year} {2022})}\BibitemShut {NoStop}%
\bibitem [{\citenamefont {Gompper}\ \emph {et~al.}(2020)\citenamefont {Gompper}, \citenamefont {Winkler}, \citenamefont {Speck}, \citenamefont {Solon}, \citenamefont {Nardini}, \citenamefont {Peruani}, \citenamefont {L{\"o}wen}, \citenamefont {Golestanian}, \citenamefont {Kaupp}, \citenamefont {Alvarez} \emph {et~al.}}]{Gompper2020}%
  \BibitemOpen
  \bibfield  {author} {\bibinfo {author} {\bibfnamefont {G.}~\bibnamefont {Gompper}}, \bibinfo {author} {\bibfnamefont {R.~G.}\ \bibnamefont {Winkler}}, \bibinfo {author} {\bibfnamefont {T.}~\bibnamefont {Speck}}, \bibinfo {author} {\bibfnamefont {A.}~\bibnamefont {Solon}}, \bibinfo {author} {\bibfnamefont {C.}~\bibnamefont {Nardini}}, \bibinfo {author} {\bibfnamefont {F.}~\bibnamefont {Peruani}}, \bibinfo {author} {\bibfnamefont {H.}~\bibnamefont {L{\"o}wen}}, \bibinfo {author} {\bibfnamefont {R.}~\bibnamefont {Golestanian}}, \bibinfo {author} {\bibfnamefont {U.~B.}\ \bibnamefont {Kaupp}}, \bibinfo {author} {\bibfnamefont {L.}~\bibnamefont {Alvarez}}, \emph {et~al.},\ }\bibfield  {title} {\bibinfo {title} {The 2020 motile active matter roadmap},\ }\href@noop {} {\bibfield  {journal} {\bibinfo  {journal} {J. Phys - Condens. Mat.}\ }\textbf {\bibinfo {volume} {32}},\ \bibinfo {pages} {193001} (\bibinfo {year} {2020})}\BibitemShut {NoStop}%
\bibitem [{\citenamefont {Wensink}\ \emph {et~al.}(2012)\citenamefont {Wensink}, \citenamefont {Dunkel}, \citenamefont {Heidenreich}, \citenamefont {Drescher}, \citenamefont {Goldstein},\ and\ \citenamefont {L{\"o}wen}}]{Wensink2012}%
  \BibitemOpen
  \bibfield  {author} {\bibinfo {author} {\bibfnamefont {H.~H.}\ \bibnamefont {Wensink}}, \bibinfo {author} {\bibfnamefont {J.}~\bibnamefont {Dunkel}}, \bibinfo {author} {\bibfnamefont {S.}~\bibnamefont {Heidenreich}}, \bibinfo {author} {\bibfnamefont {K.}~\bibnamefont {Drescher}}, \bibinfo {author} {\bibfnamefont {R.~E.}\ \bibnamefont {Goldstein}},\ and\ \bibinfo {author} {\bibfnamefont {H.}~\bibnamefont {L{\"o}wen}},\ }\bibfield  {title} {\bibinfo {title} {Meso-scale turbulence in living fluids},\ }\href@noop {} {\bibfield  {journal} {\bibinfo  {journal} {Proceedings of the National Academy of Sciences}\ }\textbf {\bibinfo {volume} {109}},\ \bibinfo {pages} {14308} (\bibinfo {year} {2012})}\BibitemShut {NoStop}%
\bibitem [{\citenamefont {Shendruk}\ \emph {et~al.}(2017)\citenamefont {Shendruk}, \citenamefont {Doostmohammadi}, \citenamefont {Thijssen},\ and\ \citenamefont {Yeomans}}]{Shendruk2017}%
  \BibitemOpen
  \bibfield  {author} {\bibinfo {author} {\bibfnamefont {T.~N.}\ \bibnamefont {Shendruk}}, \bibinfo {author} {\bibfnamefont {A.}~\bibnamefont {Doostmohammadi}}, \bibinfo {author} {\bibfnamefont {K.}~\bibnamefont {Thijssen}},\ and\ \bibinfo {author} {\bibfnamefont {J.~M.}\ \bibnamefont {Yeomans}},\ }\bibfield  {title} {\bibinfo {title} {Dancing disclinations in confined active nematics},\ }\href@noop {} {\bibfield  {journal} {\bibinfo  {journal} {Soft Matter}\ }\textbf {\bibinfo {volume} {13}},\ \bibinfo {pages} {3853} (\bibinfo {year} {2017})}\BibitemShut {NoStop}%
\bibitem [{\citenamefont {Simha}\ and\ \citenamefont {Ramaswamy}(2002)}]{Ramaswamy2002}%
  \BibitemOpen
  \bibfield  {author} {\bibinfo {author} {\bibfnamefont {R.~A.}\ \bibnamefont {Simha}}\ and\ \bibinfo {author} {\bibfnamefont {S.}~\bibnamefont {Ramaswamy}},\ }\bibfield  {title} {\bibinfo {title} {Hydrodynamic fluctuations and instabilities in ordered suspensions of self-propelled particles},\ }\href@noop {} {\bibfield  {journal} {\bibinfo  {journal} {Phys. Rev. Lett.}\ }\textbf {\bibinfo {volume} {89}},\ \bibinfo {pages} {058101} (\bibinfo {year} {2002})}\BibitemShut {NoStop}%
\bibitem [{\citenamefont {Voituriez}\ \emph {et~al.}(2005)\citenamefont {Voituriez}, \citenamefont {Joanny},\ and\ \citenamefont {Prost}}]{Voituriez2005}%
  \BibitemOpen
  \bibfield  {author} {\bibinfo {author} {\bibfnamefont {R.}~\bibnamefont {Voituriez}}, \bibinfo {author} {\bibfnamefont {J.~F.}\ \bibnamefont {Joanny}},\ and\ \bibinfo {author} {\bibfnamefont {J.}~\bibnamefont {Prost}},\ }\bibfield  {title} {\bibinfo {title} {Spontaneous flow transition in active polar gels},\ }\href@noop {} {\bibfield  {journal} {\bibinfo  {journal} {Europhys. Lett.}\ }\textbf {\bibinfo {volume} {70}},\ \bibinfo {pages} {404} (\bibinfo {year} {2005})}\BibitemShut {NoStop}%
\bibitem [{\citenamefont {Thampi}(2022)}]{Thampi2022}%
  \BibitemOpen
  \bibfield  {author} {\bibinfo {author} {\bibfnamefont {S.~P.}\ \bibnamefont {Thampi}},\ }\bibfield  {title} {\bibinfo {title} {Channel confined active nematics},\ }\href@noop {} {\bibfield  {journal} {\bibinfo  {journal} {Current Opinion in Colloid and Interface Science}\ }\textbf {\bibinfo {volume} {61}},\ \bibinfo {pages} {101613} (\bibinfo {year} {2022})}\BibitemShut {NoStop}%
\bibitem [{\citenamefont {Opathalage}\ \emph {et~al.}(2019)\citenamefont {Opathalage}, \citenamefont {Norton}, \citenamefont {Juniper}, \citenamefont {Langeslay}, \citenamefont {Aghvami}, \citenamefont {Fraden},\ and\ \citenamefont {Dogic}}]{Opathalage2019}%
  \BibitemOpen
  \bibfield  {author} {\bibinfo {author} {\bibfnamefont {A.}~\bibnamefont {Opathalage}}, \bibinfo {author} {\bibfnamefont {M.~M.}\ \bibnamefont {Norton}}, \bibinfo {author} {\bibfnamefont {M.~P.~N.}\ \bibnamefont {Juniper}}, \bibinfo {author} {\bibfnamefont {B.}~\bibnamefont {Langeslay}}, \bibinfo {author} {\bibfnamefont {S.~A.}\ \bibnamefont {Aghvami}}, \bibinfo {author} {\bibfnamefont {S.}~\bibnamefont {Fraden}},\ and\ \bibinfo {author} {\bibfnamefont {Z.}~\bibnamefont {Dogic}},\ }\bibfield  {title} {\bibinfo {title} {Self-organized dynamics and the transition to turbulence of confined active nematics},\ }\href@noop {} {\bibfield  {journal} {\bibinfo  {journal} {Proceedings of the National Academy of Sciences}\ }\textbf {\bibinfo {volume} {116}},\ \bibinfo {pages} {4788} (\bibinfo {year} {2019})}\BibitemShut {NoStop}%
\bibitem [{\citenamefont {Hardouin}\ \emph {et~al.}(2019)\citenamefont {Hardouin}, \citenamefont {Hughes}, \citenamefont {Doostmohammadi}, \citenamefont {Laurent}, \citenamefont {Lopez-Leon}, \citenamefont {Yeomans}, \citenamefont {Ignes-Mullol},\ and\ \citenamefont {Sagues}}]{Hardouin2019}%
  \BibitemOpen
  \bibfield  {author} {\bibinfo {author} {\bibfnamefont {J.}~\bibnamefont {Hardouin}}, \bibinfo {author} {\bibfnamefont {R.}~\bibnamefont {Hughes}}, \bibinfo {author} {\bibfnamefont {A.}~\bibnamefont {Doostmohammadi}}, \bibinfo {author} {\bibfnamefont {J.}~\bibnamefont {Laurent}}, \bibinfo {author} {\bibfnamefont {T.}~\bibnamefont {Lopez-Leon}}, \bibinfo {author} {\bibfnamefont {J.~M.}\ \bibnamefont {Yeomans}}, \bibinfo {author} {\bibfnamefont {J.}~\bibnamefont {Ignes-Mullol}},\ and\ \bibinfo {author} {\bibfnamefont {F.}~\bibnamefont {Sagues}},\ }\bibfield  {title} {\bibinfo {title} {Reconfigurable flows and defect landscape of confined active nematics},\ }\href@noop {} {\bibfield  {journal} {\bibinfo  {journal} {Communications Physics}\ }\textbf {\bibinfo {volume} {2}},\ \bibinfo {pages} {121} (\bibinfo {year} {2019})}\BibitemShut {NoStop}%
\bibitem [{\citenamefont {Samui}\ \emph {et~al.}(2021)\citenamefont {Samui}, \citenamefont {Yeomans},\ and\ \citenamefont {Thampi}}]{Samui2021}%
  \BibitemOpen
  \bibfield  {author} {\bibinfo {author} {\bibfnamefont {A.}~\bibnamefont {Samui}}, \bibinfo {author} {\bibfnamefont {J.~M.}\ \bibnamefont {Yeomans}},\ and\ \bibinfo {author} {\bibfnamefont {S.~P.}\ \bibnamefont {Thampi}},\ }\bibfield  {title} {\bibinfo {title} {Flow transitions and length scales of a channel-confined active nematic},\ }\href@noop {} {\bibfield  {journal} {\bibinfo  {journal} {Soft Matter}\ }\textbf {\bibinfo {volume} {17}},\ \bibinfo {pages} {10640} (\bibinfo {year} {2021})}\BibitemShut {NoStop}%
\bibitem [{\citenamefont {Sousa}\ \emph {et~al.}(2026)\citenamefont {Sousa}, \citenamefont {Thijssen},\ and\ \citenamefont {Doostmohammadi}}]{sousa2026spontaneous}%
  \BibitemOpen
  \bibfield  {author} {\bibinfo {author} {\bibfnamefont {N.~d.~G.}\ \bibnamefont {Sousa}}, \bibinfo {author} {\bibfnamefont {K.}~\bibnamefont {Thijssen}},\ and\ \bibinfo {author} {\bibfnamefont {A.}~\bibnamefont {Doostmohammadi}},\ }\bibfield  {title} {\bibinfo {title} {Spontaneous vortex instability in active nematics},\ }\href@noop {} {\bibfield  {journal} {\bibinfo  {journal} {arXiv preprint arXiv:2609.15093}\ } (\bibinfo {year} {2026})}\BibitemShut {NoStop}%
\bibitem [{\citenamefont {Zienkiewicz}\ \emph {et~al.}(2013)\citenamefont {Zienkiewicz}, \citenamefont {Taylor},\ and\ \citenamefont {Zhu}}]{zienkiewicz2013}%
  \BibitemOpen
  \bibfield  {author} {\bibinfo {author} {\bibfnamefont {O.~C.}\ \bibnamefont {Zienkiewicz}}, \bibinfo {author} {\bibfnamefont {R.~L.}\ \bibnamefont {Taylor}},\ and\ \bibinfo {author} {\bibfnamefont {J.~Z.}\ \bibnamefont {Zhu}},\ }\href@noop {} {\emph {\bibinfo {title} {The Finite Element Method: Its Basis and Fundamentals}}},\ \bibinfo {edition} {7th}\ ed.\ (\bibinfo  {publisher} {Butterworth-Heinemann},\ \bibinfo {address} {Oxford},\ \bibinfo {year} {2013})\BibitemShut {NoStop}%
\bibitem [{\citenamefont {Beris}\ and\ \citenamefont {Edwards}(1994)}]{Beris1994}%
  \BibitemOpen
  \bibfield  {author} {\bibinfo {author} {\bibfnamefont {A.~N.}\ \bibnamefont {Beris}}\ and\ \bibinfo {author} {\bibfnamefont {B.~J.}\ \bibnamefont {Edwards}},\ }\href@noop {} {\emph {\bibinfo {title} {Thermodynamics of Flowing Systems: With Internal Microstructure}}}\ (\bibinfo  {publisher} {Oxford University Press},\ \bibinfo {year} {1994})\BibitemShut {NoStop}%
\bibitem [{\citenamefont {Lorenz}(1963)}]{Lorenz1963}%
  \BibitemOpen
  \bibfield  {author} {\bibinfo {author} {\bibfnamefont {E.~N.}\ \bibnamefont {Lorenz}},\ }\bibfield  {title} {\bibinfo {title} {{Deterministic nonperiodic flow}},\ }\href@noop {} {\bibfield  {journal} {\bibinfo  {journal} {J. Atmos. Sci.}\ }\textbf {\bibinfo {volume} {20}},\ \bibinfo {pages} {130} (\bibinfo {year} {1963})}\BibitemShut {NoStop}%
\bibitem [{\citenamefont {Das}\ \emph {et~al.}(2020)\citenamefont {Das}, \citenamefont {Bhattacharjee},\ and\ \citenamefont {Kirkpatrick}}]{Das2020}%
  \BibitemOpen
  \bibfield  {author} {\bibinfo {author} {\bibfnamefont {A.}~\bibnamefont {Das}}, \bibinfo {author} {\bibfnamefont {J.~K.}\ \bibnamefont {Bhattacharjee}},\ and\ \bibinfo {author} {\bibfnamefont {T.~R.}\ \bibnamefont {Kirkpatrick}},\ }\bibfield  {title} {\bibinfo {title} {Transition to turbulence in driven active matter},\ }\href@noop {} {\bibfield  {journal} {\bibinfo  {journal} {Phys. Rev. E}\ }\textbf {\bibinfo {volume} {101}},\ \bibinfo {pages} {023103} (\bibinfo {year} {2020})}\BibitemShut {NoStop}%
\bibitem [{\citenamefont {Wu}\ \emph {et~al.}(2017)\citenamefont {Wu}, \citenamefont {Hishamunda}, \citenamefont {Chen}, \citenamefont {DeCamp}, \citenamefont {Chang}, \citenamefont {Fern{\'a}ndez-Nieves}, \citenamefont {Fraden},\ and\ \citenamefont {Dogic}}]{wu2017transition}%
  \BibitemOpen
  \bibfield  {author} {\bibinfo {author} {\bibfnamefont {K.-T.}\ \bibnamefont {Wu}}, \bibinfo {author} {\bibfnamefont {J.~B.}\ \bibnamefont {Hishamunda}}, \bibinfo {author} {\bibfnamefont {D.~T.}\ \bibnamefont {Chen}}, \bibinfo {author} {\bibfnamefont {S.~J.}\ \bibnamefont {DeCamp}}, \bibinfo {author} {\bibfnamefont {Y.-W.}\ \bibnamefont {Chang}}, \bibinfo {author} {\bibfnamefont {A.}~\bibnamefont {Fern{\'a}ndez-Nieves}}, \bibinfo {author} {\bibfnamefont {S.}~\bibnamefont {Fraden}},\ and\ \bibinfo {author} {\bibfnamefont {Z.}~\bibnamefont {Dogic}},\ }\bibfield  {title} {\bibinfo {title} {Transition from turbulent to coherent flows in confined three-dimensional active fluids},\ }\href@noop {} {\bibfield  {journal} {\bibinfo  {journal} {Science}\ }\textbf {\bibinfo {volume} {355}},\ \bibinfo {pages} {eaal1979} (\bibinfo {year} {2017})}\BibitemShut {NoStop}%
\bibitem [{\citenamefont {Giomi}(2015)}]{giomi2015geometry}%
  \BibitemOpen
  \bibfield  {author} {\bibinfo {author} {\bibfnamefont {L.}~\bibnamefont {Giomi}},\ }\bibfield  {title} {\bibinfo {title} {Geometry and topology of turbulence in active nematics},\ }\href@noop {} {\bibfield  {journal} {\bibinfo  {journal} {Physical Review X}\ }\textbf {\bibinfo {volume} {5}},\ \bibinfo {pages} {031003} (\bibinfo {year} {2015})}\BibitemShut {NoStop}%
\bibitem [{\citenamefont {Blanch-Mercader}\ \emph {et~al.}(2018)\citenamefont {Blanch-Mercader}, \citenamefont {Yashunsky}, \citenamefont {Garcia}, \citenamefont {Duclos}, \citenamefont {Giomi},\ and\ \citenamefont {Silberzan}}]{blanch2018turbulent}%
  \BibitemOpen
  \bibfield  {author} {\bibinfo {author} {\bibfnamefont {C.}~\bibnamefont {Blanch-Mercader}}, \bibinfo {author} {\bibfnamefont {V.}~\bibnamefont {Yashunsky}}, \bibinfo {author} {\bibfnamefont {S.}~\bibnamefont {Garcia}}, \bibinfo {author} {\bibfnamefont {G.}~\bibnamefont {Duclos}}, \bibinfo {author} {\bibfnamefont {L.}~\bibnamefont {Giomi}},\ and\ \bibinfo {author} {\bibfnamefont {P.}~\bibnamefont {Silberzan}},\ }\bibfield  {title} {\bibinfo {title} {Turbulent dynamics of epithelial cell cultures},\ }\href@noop {} {\bibfield  {journal} {\bibinfo  {journal} {Phys. Rev. Lett.}\ }\textbf {\bibinfo {volume} {120}},\ \bibinfo {pages} {208101} (\bibinfo {year} {2018})}\BibitemShut {NoStop}%
\end{thebibliography}%

\section*{End Matter}
\setcounter{equation}{0}
\renewcommand{\theequation}{S\arabic{equation}}
\paragraph*{Quiescent base states ---}
The homogeneous quiescent states are given by Eq.~\eqref{eq:base}.
To interpret the two ordered branches, we write the two-dimensional nematic
tensor in terms of the scalar order parameter $S$ and director angle $\theta$,
\begin{equation}
Q_{xx}=\frac{S}{2}\cos(2\theta),
\qquad
Q_{xy}=\frac{S}{2}\sin(2\theta).
\label{eq:end_director}
\end{equation}
Therefore, the branches
$A^*=+\frac{S_0}{2}$ and $A^*=-\frac{S_0}{2}$ respectively correspond to $\theta=0$, with the director parallel, and $\theta=\pi/2$, with the director perpendicular, to the $x-$ axis. 

\paragraph*{Transverse one-mode stability ---}


We introduce the perturbations described by Eq.~\eqref{eq:ansatz}. Note that velocity perturbations are incompressible since they are introduced through stream function.
It is convenient to define $\Lambda=q^2+1$ and $\Delta=q^2-1$.
for which
\begin{equation}
\nabla^2f=-\Lambda f,
\qquad
\nabla^4f=\Lambda^2f.
\label{eq:end_laplacian}
\end{equation}

\paragraph*{Stokes slaving relation ---}

Taking the curl of
the Stokes equation,
\begin{equation}
0=-\nabla p+\nabla^2\mathbf{u}
-\alpha\nabla\cdot\mathbf{Q},
\end{equation}
eliminates the pressure and gives
\begin{equation}
\nabla^4\psi
=
\alpha\left[
(\partial_y^2-\partial_x^2)Q_{xy}
+2\partial_x\partial_yQ_{xx}
\right].
\label{eq:end_curlstokes}
\end{equation}
For the perturbation in Eq.~\eqref{eq:ansatz},
$Q_{xx}=A^*$ remains spatially uniform, and hence
\begin{equation}
\partial_x\partial_yQ_{xx}=0.
\end{equation}
Furthermore,
\begin{equation}
(\partial_y^2-\partial_x^2)\delta Q_{xy}
=
(q^2-1)Yf
=
\Delta Yf.
\end{equation}
Equation~\eqref{eq:end_curlstokes} therefore reduces to
\begin{equation}
\Lambda^2Xf=\alpha\Delta Yf.
\end{equation}
Since both sides are proportional to the same Fourier mode, projection onto
$f$ gives Eq.~\eqref{eq:slaving}.

\paragraph*{Linearized nematic dynamics --}

We next project the $xy$ component of the Beris--Edwards equation,
\begin{align}
\partial_tQ_{xy}
+\mathbf{u}\cdot\nabla Q_{xy}
={}&
S_{xy}
+K\nabla^2Q_{xy}
\nonumber\\
&+
C\left(
S_0^2-4Q_{xx}^2-4Q_{xy}^2
\right)Q_{xy},
\label{eq:end_Qxy}
\end{align}
onto the same Fourier mode. The perturbation velocity generated by
Eq.~\eqref{eq:ansatz} is
\begin{equation}
u_x=-X\cos(qx)\sin y,
\qquad
u_y=qX\sin(qx)\cos y.
\label{eq:end_velocity}
\end{equation}
Since the base state is spatially uniform, advection contributes only at
quadratic order in the perturbation amplitudes. Indeed, for the present
one-mode ansatz,
\begin{equation}
\mathbf{u}\cdot\nabla\delta Q_{xy}=0
\label{eq:end_advection}
\end{equation}
pointwise. The relevant strain-rate and vorticity components are $E_{xy}
=
\frac{\Delta}{2}Xf,$ and $\Omega_{xy}
=
-\frac{\Lambda}{2}Xf.$
The diagonal strain component is
\begin{equation}
E_{xx}
=
\partial_xu_x
=
qX\sin(qx)\sin y.
\label{eq:end_Exx}
\end{equation}

For a symmetric and traceless two-dimensional nematic tensor, the
off-diagonal Beris--Edwards co-rotation term is
\begin{equation}
S_{xy}
=
\lambda E_{xy}
-2Q_{xx}\Omega_{xy}
-4\lambda Q_{xx}Q_{xy}E_{xx}
-4\lambda Q_{xy}^2E_{xy}.
\label{eq:end_Sxy_general}
\end{equation}
At linear order in $(X,Y)$, the first two terms give
\begin{align}
\lambda E_{xy}-2A^*\Omega_{xy}
&=
\left(
\frac{\lambda}{2}\Delta
+A^*\Lambda
\right)Xf.
\label{eq:end_Sxy_linear}
\end{align}

The third term in Eq.~\eqref{eq:end_Sxy_general} is quadratic in the
perturbation amplitudes. It can also be verified that this contribution is orthogonal to the retained Fourier mode.
The final
term in Eq.~\eqref{eq:end_Sxy_general} is cubic and therefore does not enter
the linear stability calculation. Consequently,
\begin{equation}
\mathcal{P}_f[S_{xy}]
=
\left(
A^*\Lambda+\frac{\lambda}{2}\Delta
\right)X
+\mathcal{O}(Y^2),
\label{eq:end_Sxy_projected_linear}
\end{equation}
where
\begin{equation}
\mathcal{P}_f[g]
=
\frac{\langle fg\rangle}{\langle f^2\rangle}
\end{equation}
denotes projection onto the retained mode.

The elastic contribution is
\begin{equation}
\mathcal{P}_f[K\nabla^2Q_{xy}]
=
-K\Lambda Y.
\label{eq:end_elastic}
\end{equation}
Linearizing the bulk term about $Q_{xx}=A^*$ and $Q_{xy}=0$ gives
\begin{equation}
\mathcal{P}_f[\mathrm{bulk}]
=
C\left(
S_0^2-4A^{*2}
\right)Y.
\label{eq:end_bulk_linear}
\end{equation}
Combining these contributions yields
\begin{equation}
\dot Y
=
\left(
A^*\Lambda+\frac{\lambda}{2}\Delta
\right)X
+
\left[
C\left(S_0^2-4A^{*2}\right)
-K\Lambda
\right]Y.
\label{eq:end_Y_linear}
\end{equation}

Finally, eliminating $X$ using the Stokes slaving relation,
Eq.~\eqref{eq:slaving}, we obtain Eq.~\eqref{eq:ydot}.


\end{document}